# Analytical Models of the Performance of C-V2X Mode 4 Vehicular Communications

Manuel Gonzalez-Martín, Miguel Sepulcre, Rafael Molina-Masegosa, Javier Gozalvez

*Abstract*— The C-V2X or LTE-V standard has been designed to support V2X (Vehicle to Everything) communications. The standard is an evolution of LTE, and it has been published by the 3GPP in Release 14. This new standard introduces the C-V2X or LTE-V Mode 4 that is specifically designed for V2V communications using the PC5 sidelink interface without any cellular infrastructure support. In Mode 4, vehicles autonomously select and manage their radio resources. Mode 4 is highly relevant since V2V safety applications cannot depend on the availability of infrastructure-based cellular coverage. This paper presents the first analytical models of the communication performance of C-V2X or LTE-V Mode 4. In particular, the paper presents analytical models for the average PDR (Packet Delivery Ratio) as a function of the distance between transmitter and receiver, and for the four different types of transmission errors that can be encountered in C-V2X Mode 4. The models are validated for a wide range of transmission parameters and traffic densities. To this aim, this study compares the results obtained with the analytical models to those obtained with a C-V2X Mode 4 simulator implemented over Veins.

## I. Introduction

Vehicular networks are essential to support active traffic safety and advanced management applications [1]. The Third Generation Partnership Project (3GPP) published in Release 14 an evolution of the LTE standard to support V2X (Vehicle to Everything) communications. This evolution is commonly referred to as C-V2X, Cellular V2X, LTE-V, LTE-V2X or LTE-V2V [2]. C-V2X is considered an alternative to IEEE 802.11p since it supports direct communication between vehicles using the PC5 interface (also known as V2X sidelink communications). Release 14 introduces two new communication modes (Mode 3 and Mode 4) specifically designed for V2V (Vehicle to Vehicle) communications and that significantly differ from Modes 1 and 2 defined in Release 12 for D2D (Device-to-Device) communications. In Mode 3, the cellular network selects and manages the radio resources used by vehicles for their direct V2V communications. In Mode 4, vehicles autonomously select and manage their radio resources without any cellular infrastructure support. To this aim, Mode 4 defines a sensing-based Semi-Persistent Scheduling (SPS) scheme that vehicles must implement to autonomously select their radio resources without the assistance of the cellular infrastructure. Mode 4 is highly relevant since V2V safety applications cannot depend on the availability of infrastructure-based cellular coverage.

Recent studies have analyzed the performance of C-V2X Mode 4, and compared it to that achieved with IEEE 802.11p standards such as DSRC or ITS-G5 [3][4]. These studies are based on network simulations, and to the authors' knowledge there are no analytical models of the C-V2X Mode 4 communication performance in the literature. Existing and recent C-V2X analytical models focus on C-V2X Mode 3 where the radio resources are managed and assigned by the infrastructure. For example, [5] proposes analytical models using combined Markov chains to evaluate the performance of different scheduling schemes in C-V2X. [6] analytically models C-V2X Mode 3, and compares its scalability to that of IEEE 802.11p. The authors utilize the model proposed in [7] to analyze the beaconing resource occupation. Prior to [5] and [6], other studies have reported analytical models for V2I (Vehicle to Infrastructure) communications using LTE. For example, [8] proposes a M/M/m queuing model to evaluate the probability that a vehicle finds all channels busy, and to derive the expected waiting times. An analytical framework is proposed in [9] to compare IEEE 802.11p and LTE in terms of the probability to deliver a packet before a deadline. This study considers that vehicles transmit their packets in an uplink channel to the LTE base station, and the base station retransmits the relevant packets to each vehicle over a downlink channel.

Analytical models are an important evaluation tool to provide information about the performance under a wide range of parameters and conditions. Analytical studies can then be complemented by more comprehensive, but also more computationally expensive, network simulations. In this context, this paper presents and validates the first analytical models of the communication performance of C-V2X Mode 4. The models provide the average PDR (Packet Delivery Ratio) as a function of the distance between transmitter and receiver. In addition, the models quantify the four different types of This work was supported in part by the *Conselleria de Educación, Investigación, Cultura y Deporte* of *Generalitat Valenciana* through the project AICO/2018/A/095, the Spanish Ministry of Economy and Competitiveness and FEDER funds under the projects TEC2014-57146-R and TEC2017-88612-R, and research grant PEJ-2014-A33622. Manuel Gonzalez-Martin, Miguel Sepulcre, Rafael Molina-Masegosa and Javier Gozalvez are with the UWICORE Laboratory, Universidad Miguel Hernandez de Elche (UMH), Spain. E-mail: magmartin10@gmail.com, msepulcre@umh.es, rafael.molinam@umh.es, j.gozalvez@umh.es.

packet errors that affect C-V2X Mode 4 [10]: errors due to half-duplex transmissions, errors due to a received signal power below the sensing power threshold, errors due to propagation effects, and errors due to packet collisions. The accuracy of the proposed models is validated by comparing their results to those obtained using a comprehensive C-V2X Mode 4 network simulator developed over Veins and presented in [10]. The model is validated for a wide range of transmission parameters and traffic densities. In particular, the model has been validated for several transmission power levels, Modulation and Coding Schemes (MCS) and sub-channelizations, and packet transmission frequencies.

## II. C-V2X MODE 4

### A. Physical layer

C-V2X utilizes SC-FDMA and supports 10 and 20MHz channels. Each channel is divided into sub-frames, Resource Blocks (RBs), and sub-channels. Sub-frames are 1ms long (like the Transmission Time Interval). A RB is the smallest unit of frequency resources that can be allocated to a LTE user. It is 180kHz wide in frequency (12 sub-carriers of 15kHz). C-V2X defines sub-channels as a group of RBs in the same sub-frame. The number of RBs per sub-channel can vary. Sub-channels are used to transmit data and control information. The data is transmitted in Transport Blocks (TBs). A TB contains a full packet to be transmitted, e.g. a beacon or a CAM (Cooperative Awareness Message)/BSM (Basic Safety Message). TBs can be transmitted using QPSK or 16-QAM and turbo coding. Each TB is transmitted with a Sidelink Control Information (SCI) that occupies 2 RBs in the same sub-frame, and represents the signaling overhead in C-V2X Mode 4. The SCI includes information such as the modulation and coding scheme used to transmit the TB, and the RBs used to transmit the TB. Its correct reception is necessary for other vehicles to be able to receive and decode the transmitted TB. The maximum transmit power is 23dBm, and the standard requires a sensitivity power level at the receiver of -90.4dBm [11]. Fig. 1 (left part) illustrates the C-V2X sub-channelization when the available bandwidth is divided in 4 sub-channels and the control information is adjacent to the data.

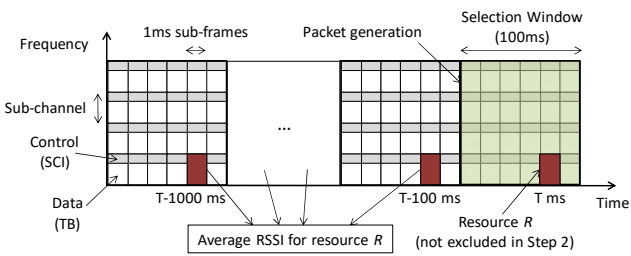

Fig. 1. C-V2X sub-channelization and average RSSI of a resource for $\lambda$=10Hz

### B. Sensing-based Semi-Persistent Scheduling

In C-V2X Mode 4, vehicles autonomously select their resources without the assistance of the cellular infrastructure. To this aim, they use the sensing-based SPS scheduling scheme specified in Release 14 [12][13]. A vehicle reserves the selected resource(s) for a random number of consecutive packets. This number depends on the number of packets transmitted per second ($\lambda$), or inversely the packet transmission interval. For $\lambda$=10Hz, 20Hz and 50Hz, this random number is selected between 5 and 15, between 10 and 30, and between 25 and 75, respectively. When a vehicle needs to reserve new resources, it randomly selects a Reselection Counter. After each transmission, the Reselection Counter is decremented by one. When it is equal to zero, new resources must be selected and reserved with probability (1-$p_{res}$) where $p_{res} \in [0,0.8]$[1]. Each vehicle includes its packet transmission interval and the value of its Reselection Counter in its SCI. Vehicles use this information to estimate which resources are free when making their own reservation to reduce packet collisions. The process to reserve resources is organized in the following 3 steps.

*Step 1.* When a vehicle $v_t$ needs to transmit a new packet and the Reselection Counter is zero, $v_t$ has to reserve new resources within a Selection Window. The Selection Window is the time window between the time the packet has been generated ($t_b$) and the defined maximum latency (Fig. 1, right part). The maximum latency is 100ms for $\lambda$=10Hz, 50ms for $\lambda$=20Hz and 20ms for $\lambda$=50Hz [12]. Within the Selection Window, the vehicle identifies the resources it could reserve. A resource is a group of adjacent sub-channels within the same sub-frame where the packet (SCI+TB) to be transmitted fits.

*Step 2.* Vehicle $v_t$ then creates a list $L_A$ of available resources it could reserve. This list includes all the resources previously identified in Step 1 except those that meet the following two conditions:

1) $v_t$ has received in the last 1000 sub-frames an SCI from another vehicle indicating that it will utilize this resource in the Selection Window or any of its next *Reselection Counter* packets.

2) $v_t$ measures an average Reference Signal Received Power (RSRP) over the resource higher than a given threshold.

Vehicle $v_t$ also excludes all the resources of sub-frame $f_i$ in the Selection Window if $v_t$ was transmitting during any previous sub-frame $f_j$, where $j=i-100 \cdot k$ and $k \in N$, $1 \leq k \leq 10$ for $\lambda$=10Hz[2].

After Step 2 is executed, $L_A$ must contain at least 20% of all the resources initially identified in the Selection Window during Step 1. If not, Step 2 is iteratively executed until the 20% target is met. In each iteration, the RSRP threshold is increased by 3dB.

*Step 3.* $v_t$ creates a list of candidate resources $L_C$ that includes the resources in $L_A$ that experienced the lowest average RSSI (Received Signal Strength Indicator). The size of $L_C$ must be equal to the 20% of all the resources in the Selection Window identified during Step 1. The RSSI value is averaged over all the previous $t_R$-100·$j$ sub-frames ($j \in N$, $1 \leq j$

---

[1] $p_{res}$ is usually set equal to 0 [14]. This value is assumed in this study.
[2] For $\lambda$=20Hz, $j=i-50 \cdot k$ and $k \in N$, $1 \leq k \leq 20$. For $\lambda$=50Hz, $j=i-20 \cdot k$ and $k \in N$, $1 \leq k \leq 50$.

≤ 10) for $\lambda$=10Hz[3] (Fig. 1). Vehicle $v_t$ then randomly chooses one of the candidate resources in $L_C$, and reserves it for the next Reselection Counter transmissions.

### III. TRANSMISSION ERRORS IN C-V2X MODE 4

C-V2X Mode 4 transmissions can encounter the following four mutually exclusive errors that are analytically quantified in the next section:

1) Errors due to half-duplex transmissions (*HD*). The C-V2X radio is half-duplex. This error is then produced when a packet cannot be received by a vehicle because the vehicle is transmitting its own packet in the same sub-frame. This type of error does not depend on the distance between transmitter and receiver, but just on the probability that two vehicles select the same sub-frame to transmit their packets. The probability of not correctly receiving a packet due to this effect is here referred to as $\delta_{HD}$:

$$\delta_{HD} = \Pr(e = HD) \qquad (1)$$

2) Error due to a received signal power below the sensing power threshold (*SEN*). This error is produced when a packet is received with a signal power below the sensing power threshold $P_{SEN}$, and hence it cannot be decoded. This error mainly depends on the transmission power, the sensing power threshold, the propagation and the distance between transmitter and receiver. This type of error excludes those quantified in 1). The probability of not correctly receiving a packet due to this effect is here referred to as $\delta_{SEN}$:

$$\delta_{SEN} = \Pr(e = SEN \mid e \neq HD) \qquad (2)$$

3) Error due to propagation effects. In this case, a packet is received with a signal power higher than $P_{SEN}$, but the received SNR (Signal to Noise Ratio) is not sufficient to guarantee the correct reception of the packet. This type of error does not consider interferences and collisions (i.e. it is only due to propagation), and hence depends on the same factors as 2) plus also on the MCS. In this study, this type of error excludes those quantified in 1) and 2). The probability of not correctly receiving a packet due to this effect is here referred to as $\delta_{PRO}$:

$$\delta_{PRO} = \Pr(e = PRO \mid e \neq HD, e \neq SEN) \qquad (3)$$

4) Error due to packet collisions (*COL*). This error is produced when a vehicle transmits on the same resource (i.e. the same sub-channel and sub-frame) than another vehicle, and the interference generated prevents the correct reception of the packet by the receiver due to insufficient SINR (Signal to Interference and Noise Ratio). It depends on the configuration and operation of the SPS scheme of C-V2X Mode 4, as well as on the transmission parameters, the propagation, distance between transmitter and receiver and traffic density. In study, this type of error excludes those quantified in 1), 2) and 3). The probability of not correctly receiving a packet due to this effect is referred to as $\delta_{COL}$:

$$\delta_{COL} = \Pr(e = COL \mid e \neq HD, e \neq SEN, e \neq PRO) \qquad (4)$$

### IV. ANALYTICAL MODELS

This section analytically quantifies the four possible transmission errors in C-V2X Mode 4, and derives an analytical model of the PDR as a function of the distance between transmitter and receiver. To this aim, we consider a highway scenario with multiple lanes where vehicles are separated by $1/\beta$ meters (i.e. a traffic density of $\beta$ vehicles per meter). All vehicles periodically transmit $\lambda$ packets per second on the same 10MHz channel with transmission power $P_t$. We consider that all packets have the same size (*B* bytes) and are transmitted using the same MCS.

To derive the analytical expressions, we consider that vehicle $v_t$ will act as transmitter, and vehicle $v_r$ as receiver. Both vehicles are separated by a distance $d_{t,r}$. The analytical models proposed consider that a packet is correctly received if none of the identified types of error occur. Since these errors are exclusive, the PDR can be calculated as:

$$PDR(d_{t,r}) = (1-\delta_{HD}) \cdot (1-\delta_{SEN}(d_{t,r})) \\ \cdot (1-\delta_{PRO}(d_{t,r})) \cdot (1-\delta_{COL}(d_{t,r})) \qquad (5)$$

We can normalize the probability of each type of error, and express the PDR as:

$$PDR = 1 - \hat{\delta}_{HD} - \hat{\delta}_{SEN} - \hat{\delta}_{PRO} - \hat{\delta}_{COL} \qquad (6)$$

where

$$\hat{\delta}_{HD} = \delta_{HD} \qquad (6.1)$$

$$\hat{\delta}_{SEN}(d_{t,r}) = (1-\delta_{HD}) \cdot \delta_{SEN}(d_{t,r}) \qquad (6.2)$$

$$\hat{\delta}_{PRO}(d_{t,r}) = (1-\delta_{HD}) \cdot (1-\delta_{SEN}(d_{t,r})) \cdot \delta_{PRO}(d_{t,r}) \qquad (6.3)$$

$$\hat{\delta}_{COL}(d_{t,r}) = (1-\delta_{HD}) \cdot (1-\delta_{SEN}(d_{t,r})) \cdot (1-\delta_{PRO}(d_{t,r})) \cdot \delta_{COL}(d_{t,r}) \qquad (6.4)$$

$$0 \leq \delta_{HD}, \delta_{SEN}, \delta_{PRO}, \delta_{COL} \leq 1 \qquad (6.5)$$

$$0 \leq \hat{\delta}_{HD} + \hat{\delta}_{SEN} + \hat{\delta}_{PRO} + \hat{\delta}_{COL} \leq 1 \qquad (6.6)$$

Appendix A shows how to derive eq. (5) from eq. (6) using eq. (6.1)-(6.4). Table I identifies the sub-sections and equations used to describe each of the four types of error. This table can be used as a reference by the reader to follow the description of the analytical models. Table II also lists the variables and parameters used to derive and describe the models.

TABLE I. EQUATIONS AND SECTIONS DESCRIBING EACH TYPE OF ERROR

| Type of error | Variable | Equation(s) | Sub-section |
|---|---|---|---|
| Half duplex | $\delta_{HD}$ | (7) | IV.A |
| Sensing | $\delta_{SEN}$ | (8) to (10) | IV.B |
| Propagation | $\delta_{PRO}$ | (12) to (13.1) | IV.C |
| Collision | $\delta_{COL}$ | (14) to (34) | IV.D |

---

[3] For $\lambda$=20Hz, $t_R$·50·$j$ sub-frames ($j \in N$, 1 ≤ $j$ ≤ 20). For $\lambda$=50Hz, $t_R$·20·$j$ sub-frames ($j \in N$, 1 ≤ $j$ ≤ 50).

TABLE II. VARIABLES

| Variable | Description |
|---|---|
| $\alpha$ | Weighting factor that represents the impact of Step 2 and Step 3 in the selection of the $N_C$ candidate resources |
| $\beta$ | Traffic density (vehicles/meter) |
| CBR | Channel Busy Ratio |
| $\Delta$ | Increment of the sensing threshold (dB) |
| $\delta_{COL}$ | Probability of packet loss due to collision from any vehicle |
| $\delta_{COL}^i$ | Probability of packet loss due to collision from vehicle $v_i$ |
| $\delta_{HD}$ | Probability of packet loss due to half-duplex effect |
| $\delta_{PRO}$ | Probability of packet loss due to propagation effects |
| $\delta_{SEN}$ | Probability of packet loss due to received signal below sensing threshold |
| $\lambda$ | Number of packets transmitted per second per vehicle (Hz) |
| $N$ | Total number of resources contained in the Selection Window |
| $N_A$ | Number of assignable resources (not excluded by Step 2) |
| $N_C$ | Number of candidate resources after Steps 2 and 3 |
| $N_E$ | Number of resources excluded in Step 2 |
| $C_A(d_{t,i})$ | Number of common available resources between $v_t$ and $v_i$ |
| $C_C(d_{t,i})$ | Number of common candidate resources between $v_t$ and $v_i$ |
| $C_E(d_{t,i})$ | Number of common excluded resources between $v_t$ and $v_i$ |
| PDR | Packet Delivery Ratio |
| $P_i$ | Received interference power from vehicle $v_i$ (dBm) |
| $P_r$ | Received signal power from vehicle $v_t$ (dBm) |
| $P_{SEN}$ | Sensing threshold (dBm) |
| $P_t$ | Transmission power (dBm) |
| $p_{INT}(d_{t,i})$ | Probability that interference from $v_i$ is higher than threshold |
| $p_{SINR}(d_{t,i})$ | Probability of packet loss due to low SINR |
| $p_{SIM}(d_{t,i})$ | Probability that $v_t$ and $v_i$ simultaneously transmit using the same resource |
| $p_{SIM}^{[2]}(d_{t,i})$ | Probability that $v_t$ and $v_i$ simultaneously transmit using the same resource when only Step 2 is executed |
| $p_{SIM}^{[3]}(d_{t,i})$ | Probability that $v_t$ and $v_i$ simultaneously transmit using the same resource when only Step 3 is executed |
| $S$ | Number of resources per sub-frame |
| SINR | Signal-to-Interference-and-Noise Ratio (dB) |
| SNR | Signal-to-Noise Ratio (dB) |
| $S_{PSR}$ | Average number of vehicles that a vehicle could sense in the Selection Window if there were no packet collisions |

### A. Half-duplex errors

The probability that two vehicles cannot receive their packets because of the half-duplex effect does not depend on their distance, the C-V2X Mode 4 SPS scheme, or the channel occupancy. Two vehicles have certain probability of selecting the same sub-frame for transmitting their packets. This probability depends on the number of packets transmitted per vehicle per second, $\lambda$, and the number of sub-frames within a second. Considering 1ms sub-frames, the probability of not receiving a packet due to the half-duplex effect can be approximated by the following equation:

$$\delta_{HD} = \frac{\lambda}{1000} \qquad (7)$$

This effect is local and only affects those vehicles transmitting in the same sub-frame, i.e. vehicles transmitting in other sub-frames can still receive the packets.

### B. Errors due to a received signal power below the sensing power threshold

To calculate the probability of receiving a packet with a signal power below the sensing power threshold, we take into account the pathloss (PL) and shadowing (SH). The pathloss represents the average signal attenuation with the distance between transmitter and receiver ($d_{t,r}$) and is typically modeled with a log-distance function. The shadowing represents the effect of obstacles on the signal attenuation, and is modeled with a log-normal random distribution with zero mean and variance σ. The received signal power $P_r$ at the receiver is hence a random variable that can be expressed as:

$$P_r(d_{t,r}) = P_t - PL(d_{t,r}) - SH \qquad (8)$$

where $P_t$ is the transmission power, $PL(d_{t,r})$ is the pathloss at the distance $d_{t,r}$, and all variables are in dB. The probability that the received signal power is lower than the sensing power threshold $P_{SEN}$ is:

$$\delta_{SEN}(d_{t,r}) = \int_{-\infty}^{P_{SEN}} f_{P_r,d_{t,r}}(p)\,dp \qquad (9)$$

where $f_{P_r,d_{t,r}}(p)$ represents the PDF of the received signal power at a distance $d_{t,r}$. The shadowing follows a log-normal random distribution, so the PDF of the received signal power can be expressed as:

$$f_{P_r,d_{t,r}}(p) = \frac{1}{\sigma\sqrt{2\pi}} \exp\left(-\left(\frac{P_t - PL(d_{t,r}) - p}{\sigma\sqrt{2}}\right)^2\right) \qquad (9.1)$$

The combination of eq. (9) and (9.1) results in that the probability that the received signal power at $d_{t,r}$ is lower than the sensing power threshold is equal to:

$$\delta_{SEN}(d_{t,r}) = \frac{1}{2}\left(1 - erf\left(\frac{P_t - PL(d_{t,r}) - P_{SEN}}{\sigma\sqrt{2}}\right)\right) \qquad (10)$$

where $erf$ is the well-known error function.

$1-\delta_{SEN}$ is the PSR (Packet Sensing Ratio), and eq. (10) can be generalized to compute the PSR at any distance $d$:

$$PSR(d) = \frac{1}{2}\left(1 + erf\left(\frac{P_t - PL(d) - P_{SEN}}{\sigma\sqrt{2}}\right)\right) \qquad (11)$$

### C. Error due to propagation

The probability that a packet is lost due to propagation effects depends on the PHY layer performance of the receiver. This performance is modeled in this study using the link level performance reported in [15], and represented by means of Look-Up Tables (LUTs). These LUTs provide the Block Error Rate (BLER) as a function of the SNR for a given packet size, MCS, scenario (highway or urban), and relative speed between transmitter and receiver. To model transmission errors due to propagation effects, we consider that the SNR at a receiver is a random variable expressed in dB as:

$$SNR(d_{t,r}) = P_r(d_{t,r}) - N_0 = P_t - PL(d_{t,r}) - SH - N_0 \qquad (12)$$

where $N_0$ is the noise power. At a given distance between transmitter and receiver, $PL$ is constant, and therefore SNR follows the same random distribution as SH but with a mean value equal to $P_t - PL - N_0$. The probability that a packet is lost due to propagation effects can hence be expressed as:

$$\delta_{PRO}(d_{t,r}) = \sum_{s=-\infty}^{+\infty} BL(s) \cdot f_{SNR|P_r > P_{SEN}, d_{t,r}}(s) \qquad (13)$$

where

$$f_{SNR|P_r>P_{SEN},d_{t,r}}(s) = \begin{cases} \dfrac{f_{SNR,d_{t,r}}(s)}{1-\delta_{SEN}} & \text{if } P_r > P_{SEN} \\ 0 & \text{if } P_r \leq P_{SEN} \end{cases} \quad (13.1)$$

In eq. (13), the term $BL(s)$ represents the BLER for an SNR equal to $s$ following the LUTs in [15]. This term is multiplied by $f_{SNR|P_r>P_{SEN},d_{t,r}}(s)$, which is the PDF of the SNR experienced at a distance $d_{t,r}$ for those SNR values for which the $P_r$ is higher than $P_{SEN}$. The objective is to omit those packets with a received signal power lower than the sensing power threshold $P_{SEN}$; these packets have already been taken into account in $\delta_{SEN}$ (eq. (9)). The PDF of the $SNR$, i.e. $f_{SNR,d_{t,r}}(s)$, needs to be normalized by 1- $\delta_{SEN}$ in eq. (13.1) so that the integral of this equation between -∞ and +∞ is 1, and the probability $\delta_{PRO}$ of not receiving a packet due to propagation effects is a value between 0 and 1.

### D. Errors due to packet collisions

This error is produced when a given interfering vehicle ($v_i$) transmits on the same sub-frame and sub-channel than the transmitting vehicle ($v_t$), and the interference generated prevents the correct reception of the packet by the receiver ($v_r$). Both conditions must happen to lose a packet due to packet collision. This error depends on the link level performance, the sensing-based SPS scheduling scheme defined in C-V2X Mode 4, the scenario, and the distances between vehicles $v_t$, $v_r$ and $v_i$. Fig. 2 summarizes the steps followed to compute the probability of packet loss due to collisions ($\delta_{COL}$). This probability can be computed as a function of the probability that a vehicle $v_i$ provokes a packet loss due to collision ($\delta_{COL}^i$) with the following equation:

$$\delta_{COL}(d_{t,r}) = 1 - \prod_i \left(1 - \delta_{COL}^i(d_{t,r},d_{t,i},d_{i,r})\right) \quad (14)$$

$v_i$ can provoke a packet loss due to collision if $v_t$ and $v_i$ simultaneously transmit using the same resource, and the interference generated by $v_i$ is such that it will provoke the packet loss. The probability of packet loss due to a collision provoked by vehicle $v_i$ can then be expressed as:

$$\delta_{COL}^i(d_{t,r},d_{t,i},d_{i,r}) = p_{SIM}(d_{t,i}) \cdot p_{INT}(d_{t,r},d_{i,r}) \quad (15)$$

$p_{SIM}(d_{t,i})$ is the probability that $v_t$ and $v_i$ simultaneously transmit using the same resource. $p_{INT}(d_{t,r},d_{i,r})$ represents the probability that the interference generated by $v_i$ on the receiver $v_r$ is higher than a threshold that would provoke that if $v_t$ and $v_i$ simultaneously transmit using the same resource, then the packet cannot be correctly received at $v_r$. $p_{INT}(d_{t,r},d_{i,r})$ depends on the scenario, the link level performance and the distances between the transmitter and receiver ($d_{t,r}$) and between the interferer and the receiver ($d_{i,r}$). On the other hand, $p_{SIM}(d_{t,i})$ depends on the MAC operation and configuration (i.e. on the sensing-based SPS scheduling scheme), as well as on the propagation conditions and the distance between $v_t$ and $v_i$.

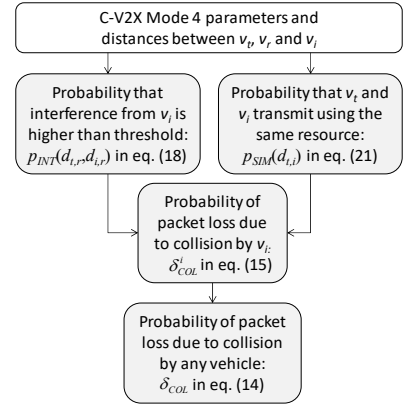

Fig. 2. Main steps to calculate the probability of packet loss due to collision.

### D1. Probability $p_{INT}(d_{t,r},d_{i,r})$ that interference is higher than threshold

To calculate $p_{INT}(d_{t,r},d_{i,r})$, we assume that the negative effect of the interference received from vehicle $v_i$ over the received signal at $v_r$ is equivalent to additional noise. The $SINR$ experienced by the receiver $v_r$ can be then expressed as:

$$SINR(d_{t,r},d_{i,r}) = P_r(d_{t,r}) - P_i(d_{i,r}) - N_0 \quad (16)$$

where all variables are in dB or dBm, and $P_i$ is the signal power received by $v_r$ from $v_i$. $SINR$ is therefore a random variable that results from the addition of two random variables ($P_r$ and $P_i$). The PDF of the $SINR$ can hence be obtained from the cross correlation of the PDF of $P_r$ and $P_i$ [16]. As a result, the probability that the receiver receives a packet with error due to low $SINR$ (i.e. low $P_r$ and/or high $P_i$) is:

$$p_{SINR}(d_{t,r},d_{i,r}) = \sum_{s=-\infty}^{+\infty} BL(s) \cdot f_{SINR|P_r>P_{SEN},d_{t,r},d_{i,r}}(s) \quad (17)$$

This equation includes the packets that could not be received due to propagation effects, i.e. those packets that would have been lost even without the interference received from $v_i$. Since these packets were already considered in $\delta_{PRO}$, we need to perform the following normalization to only consider those packets that are lost due to collisions in $p_{INT}$:

$$p_{INT}(d_{t,r},d_{i,r}) = \frac{p_{SINR}(d_{t,r},d_{i,r}) - \delta_{PRO}(d_{t,r})}{1 - \delta_{PRO}(d_{t,r})} \quad (18)$$

where $\delta_{PRO}$ is obtained from eq. (13). The same LUTs used to calculate $\delta_{PRO}$ in eq. (13) (and obtained from [15]) can be used in eq. (17) to estimate the BLER in $BL(s)$ assuming that the negative effect of the interference over the received signal is equivalent to additional noise.

### D2. Probability $p_{SIM}(d_{t,i})$ that $v_t$ and $v_i$ simultaneously transmit using the same resource

$\delta_{COL}^i$ also depends on $p_{SIM}(d_{t,i})$ as shown in eq. (15). $p_{SIM}(d_{t,i})$ represents the probability that the transmitting vehicle $v_t$ and an interfering vehicle $v_i$ transmit simultaneously in the same resource, i.e. in the same sub-channel and the same sub-frame. Fig. 3 shows the main steps needed to calculate the probability $p_{SIM}(d_{t,i})$, and that are explained next. Fig. 3 serves as a guide for the reader to follow the process.

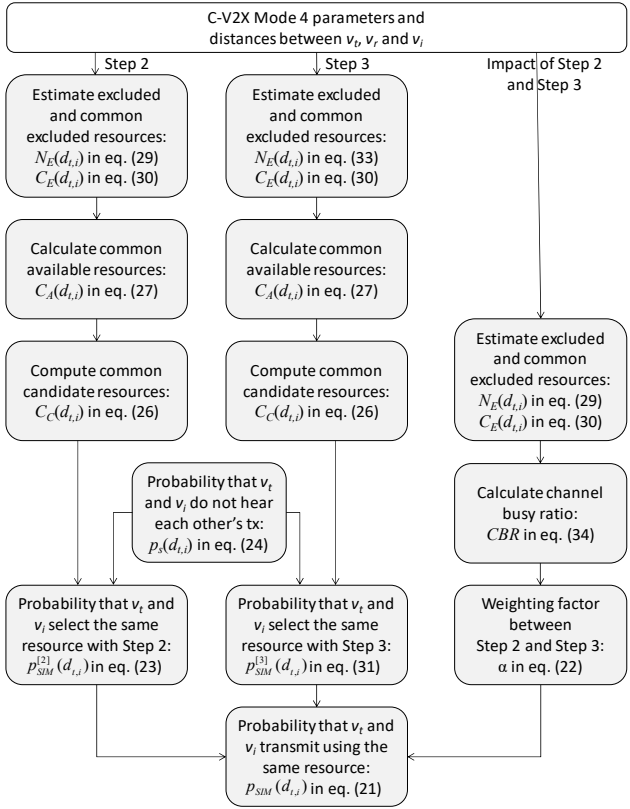

Fig. 3. Main steps followed to calculate the probability that $v_t$ and $v_i$ transmit using the same resource. The figure includes all the steps carried out while executing Steps 2 and 3 of the sensing-based SPS scheme, and the steps needed to calculate the weighting factor that represents the impact of Step 2 and Step 3 in the selection of the candidate resources.

To calculate $p_{SIM}(d_{t,i})$, we need the following definitions (see Fig. 4). $N$ is the total number of resources in all sub-frames contained in the Selection Window. $N_E$ is the number of resources excluded in Step 2 of the sensing-based SPS scheme of C-V2X Mode 4. $N_A$ is the number of assignable resources, i.e. those resources that were not excluded by Step 2 (i.e. $N_A$ is equal to the size of list $L_A$, and $N_A = N - N_E$). $N_C$ is the number of candidate resources that could be used by the transmitting vehicle after Steps 2 and 3, and is therefore equal to the size of list $L_C$. $N_C$ is equal to the 20% of $N$ [2]. Since we assume a constant traffic density, vehicles are uniformly distributed in the scenario and have the same transmission parameters, all vehicles have the same $N$, $N_E$, $N_A$ and $N_C$.

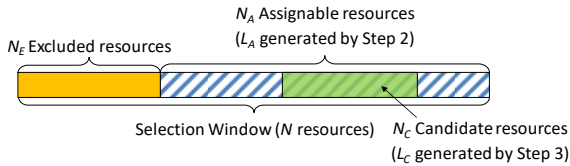

Fig. 4. Classification of resources following the sensing-based SPS scheme.

To reduce the complexity of the analytical model, we separate the derivation of $p_{SIM}(d_{t,i})$ under Steps 2 and 3 of the sensing-based SPS scheme. This approach is motivated by the fact that it is not always necessary to take into account both Steps as it is next explained:

*Step 3 has limited effect on the resource selection process when the channel load is high.* This is the case because when the channel load is high, Step 2 excludes most of the resources, and the size of the list of available resources $L_A$ is equal to the 20% of all resources in the Selection Window. Step 3 builds the list of candidate resources $L_C$ from list $L_A$. The size of $L_C$ must be equal to the 20% of all resources in the Selection Window. Thus, when the channel load is high, Step 3 will not modify the resources selected by Step 2 (Fig. 5a). As a result, when the channel load is high, we can compute $p_{SIM}(d_{t,i})$ as the probability that vehicles $v_t$ and $v_i$ transmit simultaneously in the same resource when only Step 2 is executed:

$$p_{SIM}(d_{t,i}) = p_{SIM}^{[2]}(d_{t,i}) \quad (19)$$

*Step 2 has limited effect on the resource selection process when the channel load is low.* When the channel load is low, Step 2 would exclude only a few resources to build the list of available resources $L_A$. Step 3 would build the list of candidate resources $L_C$ by selecting from $L_A$ those resources with the lowest average RSSI over the last 1000 sub-frames. Step 3 is able to exclude the resources that Step 2 would exclude, and the same $L_C$ could be obtained even if Step 2 was not executed (Fig. 5b). The utility of Step 2 is hence limited when the channel load is low. In this case, $p_{SIM}(d_{t,i})$ can be computed as the probability that $v_t$ and $v_i$ transmit simultaneously in the same resource when only Step 3 is executed:

$$p_{SIM}(d_{t,i}) = p_{SIM}^{[3]}(d_{t,i}) \quad (20)$$

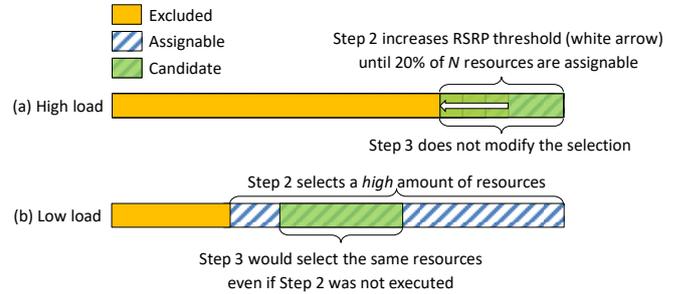

Fig. 5. Impact of Step 2 and Step 3 on $L_C$ for low and high channel loads.

*Step 2 and Step 3 need to be considered for intermediate channel load levels.* Under intermediate channel load levels, we model the probability of packet collision $p_{SIM}(d_{t,i})$ as:

$$p_{SIM}(d_{t,i}) = \alpha \cdot p_{SIM}^{[2]}(d_{t,i}) + (1-\alpha) \cdot p_{SIM}^{[3]}(d_{t,i}) \quad (21)$$

where $\alpha \in [0,1]$ is a weighting factor that represents the impact of Step 2 and Step 3 in the selection of the $N_C$ candidate resources. As previously discussed, if the channel load is high, $\alpha = 1$ because only Step 2 has an influence on the resources selected and Step 3 is not needed. If the channel load is low, $\alpha = 0$ because only Step 3 is needed. The specific value of $\alpha$ depends on the channel load, which is measured in this study using the CBR (Channel Busy Ratio) that represents the average number of resources sensed as busy. The C-V2X Mode 4 simulator described in Section V has been utilized to derive $\alpha$ through simulation. To this aim, $p_{SIM}(d_{t,i})$, $p_{SIM}^{[2]}(d_{t,i})$ and $p_{SIM}^{[3]}(d_{t,i})$ have been obtained through simulation, and their values have been used to calculate α as a function of the CBR (depicted in Fig. 6 as dots). The value of α has been derived considering a wide range of transmission parameters and traffic densities. Fig. 6 also represents the linear approximation of α that is used in our analytical model, and that is expressed as:

$$\alpha = \begin{cases} 0 & if \quad CBR < 0.2 \\ 2 \cdot CBR - 0.4 & if \quad 0.2 \leq CBR \leq 0.7 \\ 1 & if \quad CBR > 0.7 \end{cases} \quad (22)$$

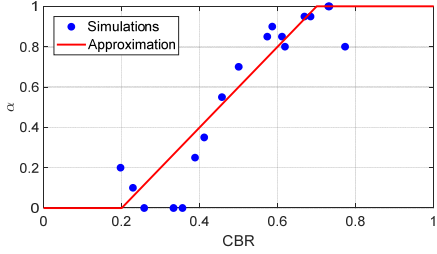

Fig. 6. Weighting factor α in eq. (21).

To derive $p_{SIM}(d_{t,i})$, we derive first $p_{SIM}^{[2]}(d_{t,i})$ and $p_{SIM}^{[3]}(d_{t,i})$, which represent the probability that vehicles $v_t$ and $v_i$ transmit using the same resource when only Step 2 or Step 3 are executed, respectively. If only Step 2 was executed, each vehicle would create its set of candidate resources $L_C$ by randomly selecting them from its set of assignable resources $L_A$ (i.e. there is no Step 3 to select the resources with lowest RSSI). Each vehicle then randomly selects the resource that will be used to transmit a packet from the set of $N_C$ candidate resources. As a result, the probability that two vehicles select the same resource for transmission depends on the number of candidate resources that they have in common. This number is here referred to as the number of common candidate resources $C_C$. Fig. 7 illustrates the concept of $C_C$. Since the $N_C$ candidate resources are selected from the $N_A$ assignable resources, $C_C$ depends on the number of common assignable resources $C_A$ (Fig. 7), i.e. on how many resources the $L_A$ lists of vehicles $v_t$ and $v_i$ have in common. In turn, $C_A$ depends on the number of common excluded resources, $C_E$ (Fig. 7). $C_E$ represents the resources excluded by both vehicles $v_t$ and $v_i$. We need to compute the number of common excluded, assignable and candidate resources ($C_E$, $C_A$ and $C_C$) for vehicles $v_t$ and $v_i$ in order to calculate the probability that $v_t$ and $v_i$ transmit using the same resource.

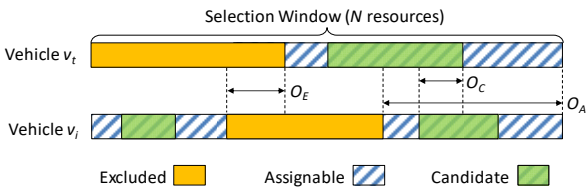

Fig. 7. Illustration of common excluded ($C_E$), assignable ($C_A$) and candidate ($C_C$) resources for two vehicles.

$p_{SIM}^{[2]}(d_{t,i})$ depends on the number of common candidate resources between vehicles $v_t$ and $v_i$, which depends on the distance between the two vehicles, $C_C(d_{t,i})$. It also depends on the probability $p_s(d_{t,i})$ that $v_t$ and $v_i$ do not take into account their respective transmissions before selecting a new resource. This can occur if the two vehicles cannot sense each other. It can also occur when $v_t$ and $v_i$ select their resources nearly at the same time, and hence they cannot take into account each other's selection as they have not been able yet to sense any packet transmitted using the newly selected resource. The probability that vehicles $v_t$ and $v_i$ transmit using the same resource when only Step 2 of the sensing-based SPS scheme is executed can then be expressed as:

$$p_{SIM}^{[2]}(d_{t,i}) = p_s(d_{t,i}) \cdot \frac{C_C(d_{t,i})}{N_C^2} \quad (23)$$

If the two vehicles are not able to take into account their respective transmissions, the probability that they simultaneously select a given resource that belongs to their set of candidate resources is $1/N_C^2$. Eq. (23) is obtained by multiplying this probability by the number of common candidate resources and the probability that they do not take into account their respective transmissions. $p_s(d_{t,i})$ depends on the probability that the transmitting and interfering vehicles ($v_t$ and $v_i$, separated by a distance $d_{t,i}$) are able to sense their respective transmissions, which is represented by the Packet Sensing Ratio $PSR(d_{t,i})$ (see eq. (11)). It also depends on the average number of consecutive packet transmissions $\tau$ for which each vehicle has to use the same resource[4]. We model the relationship between $p_s(d_{t,i})$, $PSR(d_{t,i})$, and $\tau$ as follows:

$$p_s(d_{t,i}) = 1 - (1 - 1/\tau) \cdot PSR(d_{t,i}) \quad (24)$$

If the transmitting and interfering vehicles ($v_t$ and $v_i$) are out of each other's sensing range (i.e. $PSR(d_{t,i})=0$), they will not detect their respective transmissions and hence $p_s(d_{t,i})=1$. When both vehicles are close to each other and $PSR(d_{t,i})=1$, they will detect each other and can consider their previous transmissions in the resource selection process. This is however not possible if one of the two vehicles has to select a resource, and the other vehicle has just selected its resource but has not yet made any transmission using the newly selected resource. This effect occurs with probability $1/\tau$, and therefore decreases as $\tau$ increases.

To compute $p_{SIM}^{[2]}(d_{t,i})$, we also need to calculate $C_C(d_{t,i})$ that is a function of the number of common assignable resources $C_A(d_{t,i})$. When Step 3 is not executed, $C_A(d_{t,i})$ is equal to the number of assignable resources that both the transmitting vehicle $v_t$ and the interfering vehicle $v_i$ did not exclude in Step 2 of the sensing-based SPS scheme. Since vehicles randomly select their resource from their set of assignable resources when Step 3 is not modeled, the relationship between $C_C(d_{t,i})$, $C_A(d_{t,i})$, the number of candidate resources $N_C$, and the number of assignable resources $N_A$ is:

$$\left(\frac{C_C(d_{t,i})}{N_C}\right) \cdot N_A = \left(\frac{C_A(d_{t,i})}{N_A}\right) \cdot N_C \quad (25)$$

---

[4] For example, $\tau=(15+5)/2$ in the C-V2X Mode 4 when $\lambda=10$Hz, since the Reselection Counter is randomly selected between 5 and 15.

and hence:

$$C_C(d_{t,i}) = C_A(d_{t,i}) \cdot \left(\frac{N_C}{N_A}\right)^2 \quad (26)$$

Using Fig. 7, it is possible to relate $C_A(d_{t,i})$ and $C_E(d_{t,i})$ as:

$$C_A(d_{t,i}) = N - 2 \cdot N_E + C_E(d_{t,i}) \quad (27)$$

$N$ is the total number of resources in the Selection Window, and can be computed as follows considering that there are 1000 sub-frames per second:

$$N = 1000 \cdot \frac{S}{\lambda} \quad (28)$$

where $S$ is the number of sub-channels per sub-frame, and $\lambda$ is number of packets transmitted per vehicle per second.

To compute $C_A(d_{t,i})$, we need to calculate $C_E(d_{t,i})$ and $N_E$. $N_E$ depends on the traffic density, the total number of resources in the Selection Window, the transmission power and the scenario, and is here estimated[5] as:

$$N_E = \frac{S_{PSR}}{2} + \sum_{k=1}^{S_{PSR}/2} \max\left(1 - \frac{k}{N - S_{PSR}/2}, 0\right) \quad (29)$$

where $S_{PSR}$ represents the average number of vehicles that a vehicle could sense in the Selection Window if there were no packet collisions. $S_{PSR}$ can be estimated considering that a packet transmitted by a vehicle located at a given distance $d$ is sensed if its received signal power is higher than the sensing power threshold. A vehicle located at a short distance would be sensed with probability $PSR(d)=1$, but a vehicle at a large distance will be sensed with probability $PSR(d)=0$. Vehicles at intermediate distances will be sensed with probability $0<PSR(d)<1$. $S_{PSR}$ can be then estimated as function of the packet sensing ratio with the following equation:

$$S_{PSR} = \sum_{i=-\infty}^{+\infty} PSR(d_{t,i}) = \sum_{i=-\infty}^{+\infty} PSR\left(\frac{i}{\beta}\right) = \beta \cdot \sum_{i=-\infty}^{+\infty} PSR(i) \quad (29.1)$$

where $\beta$ is the traffic density in vehicles/m. This equation considers the theory of the Riemann sum to take out the traffic density from the $PSR$ summation.

To calculate $C_E(d_{t,i})$, let's consider that a vehicle $v_k$ is transmitting in a given resource. The probability that two vehicles ($v_t$ and $v_i$) exclude the resource used by vehicle $v_k$ depends on their distance to $v_k$ ($d_{t,k}$ and $d_{i,k}$ respectively), and is equal to $PSR(d_{t,k}) \cdot PSR(d_{i,k})$. In the considered traffic scenario, this probability can also be expressed as $PSR(d_{t,i}+d_{i,k}) \cdot PSR(d_{i,k})$. If $v_t$ and $v_i$ are at the same location, they would exclude approximately the same resources because they would sense the transmissions of approximately the same vehicles. However, if vehicles $v_i$ and $v_t$ are separated by long distances, the resources excluded by each one of them can be considered independent. In this case, the proportion of common excluded resources between both vehicles tends to $N_E/N$, and therefore the number of common excluded resources $C_E$ tends to $N_E^2/N$. We can then compute $C_E$ for vehicles $v_t$ and $v_i$ separated by a distance $d_{t,i}$ as:

---
[5] This approximation has been validated through simulations using the C-V2X Mode 4 simulator presented in Section V.

$$C_E(d_{t,i}) = \frac{R_{PSR}(d_{t,i})}{R_0}\left(\frac{\beta \cdot N_E \cdot R_0}{S_{PSR}} - \frac{N_E^2}{N}\right) + \frac{N_E^2}{N} \quad (30)$$

where

$$R_0 = R_{PSR}(0) \quad (30.1)$$

and $R_{PSR}(d_{t,i})$ is the autocorrelation of the PSR function at $d_{t,i}$:

$$R_{PSR}(d_{t,i}) = \sum_{j=-\infty}^{+\infty} PSR\left(\frac{j}{\beta} + d_{t,i}\right) \cdot PSR\left(\frac{j}{\beta}\right) \quad (30.2)$$

In eq. (30.2), please note that the distance between two consecutive vehicles is $1/\beta$ when the traffic density is $\beta$, and this is why the term $j/\beta$ is introduced.

Combining eq. (23)-(30.2), we can compute $p_{SIM}^{[2]}(d_{t,i})$ that represents the probability that the transmitting vehicle $v_t$ and an interfering vehicle $v_i$ simultaneously transmit using the same resource when only Step 2 of the sensing-based SPS scheme is considered. To compute $p_{SIM}^{[3]}(d_{t,i})$, we follow a similar approach than for $p_{SIM}^{[2]}(d_{t,i})$ and can be computed as:

$$p_{SIM}^{[3]}(d_{t,i}) = p_s(d_{t,i}) \cdot \frac{C_C(d_{t,i})}{N_C^2} \quad (31)$$

The relationship between $C_C$, $C_A$ and $C_E$ is maintained whether we consider Step 2 or Step 3 of the sensing-based SPS scheme. As a result, eq. (24) to (28) and (29.1) to (30.2) obtained for Step 2 are also valid to compute $p_{SIM}^{[3]}(d_{t,i})$. This is not the case for the expression of $N_E$ that needs though to be computed when only Step 3 is executed. In this case, $L_C$ is built from the assignable resources, and Step 3 excludes the resources with the highest average RSSI experienced during the last 1000 sub-frames. When the channel load is high, it is possible that Step 3 excludes more than 80% of the resources. Since the size of $L_C$ must be equal to $0.2 \cdot N$, if more than $0.8 \cdot N$ resources are excluded, Step 3 must consider as assignable certain resources that it had previously excluded until filling $L_C$. Step 3 includes in $L_C$ the resources with the lowest average RSSI that it had previously excluded until $L_C$ is filled. This process is equivalent to increasing the sensing power threshold from $P_{SEN}$ to certain $P_{SEN}+n \cdot \Delta$ where $n$ is a positive integer and $\Delta$ is certain small increment in dB. We need to find the minimum value of $n$ that reduces the number of excluded resources to less than $0.8 \cdot N$. This is equivalent to finding the minimum value of $n$ that satisfies the following relation:

$$\frac{S_{PSR}^{(n)}}{2} + \sum_{k=1}^{S_{PSR}^{(n)}/2} \max\left(1 - \frac{k}{N - S_{PSR}^{(n)}/2}, 0\right) \leq 0.8 \cdot N \quad (32)$$

where

$$S_{PSR}^{(n)} = \sum_{i=-\infty}^{+\infty} PSR_n\left(\frac{i}{2\beta}\right) = 2\beta \cdot \sum_{i=-\infty}^{+\infty} PSR_n(i) \quad (32.1)$$

$$PSR_n(d) = \frac{1}{2}\left(1 + erf\left(\frac{P_T - PL(d) - (P_{SEN} + n \cdot \Delta)}{\sigma\sqrt{2}}\right)\right) \quad (32.2)$$

Eq. (32.1) considers a $2\beta$ factor instead of $\beta$ as in eq. (29.1). This is the case because Step 3 needs to take into account the number of different resources occupied in the last 1000 sub-frames. In 1000 sub-frames, each vehicle transmits in 2 different resources on average (i.e. it will perform one resource re-selection per second on average). For example, for $\lambda$=10Hz, each vehicle performs a resource selection every (5+15)/2=10 packet transmissions on average, i.e. every 1000 sub-frames or 1000ms. To take this effect into account in Step 3, we have estimated $S_{PSR}$ (i.e. the average number of vehicles that could be sensed if there were no packet collisions) considering that the traffic density $\beta$ is doubled. Given that the $PSR_n$ function monotonically decreases as $n$ increases, we can solve the problem by evaluating increasing values of $n$, starting at $n = 0$. Once the minimum value of $n$ that satisfies eq. (32) is found, the number of excluded resources that will not be part of $L_C$ can be approximated as:

$$N_E = \frac{S_{PSR}^{(n)}}{2} + \sum_{k=1}^{S_{PSR}^{(n)}/2} \max\left(1 - \frac{k}{N - S_{PSR}^{(n)}/2}, 0\right) \quad (33)$$

$p_{SIM}^{[3]}(d_{t,i})$ is then computed following eq. (31), and using eq. (24) to (28) and (29.1) to (30.2) and the number of excluded resources $N_E$ in eq. (33). The probability that the transmitting vehicle $v_t$ and an interfering vehicle $v_i$ simultaneously transmit using the same resource $p_{SIM}(d_{t,i})$ is then computed following eq. (21) that relates $p_{SIM}^{[2]}(d_{t,i})$ and $p_{SIM}^{[3]}(d_{t,i})$. The value of $\alpha$ is calculated using eq. (22) considering that the CBR can be analytically estimated as:

$$CBR = \frac{N_E}{N} \quad (34)$$

where the number of excluded resources $N_E$ is calculated with eq. (29) for Step 2 since it considers only those resources that are occupied in the last Selection Window.

We can then compute the probability of packet loss due to a collision provoked by vehicle $v_i$ ($\delta_{COL}^i$) using eq. (15), and eq. (18) and (21) to represent $p_{INT}(d_{t,i},d_{i,r})$ and $p_{SIM}(d_{t,i})$. The probability of packet loss due to collisions ($\delta_{COL}$) is then computed following eq. (14). Finally, the PDR is computed using eq. (5) where $\delta_{HD}$, $\delta_{SEN}$, $\delta_{PRO}$ and $\delta_{COL}$ are obtained from eq. (7), (10), (13) and (14), respectively.

## V. MODEL VALIDATION

### A. Framework and Simulation Environment

The proposed C-V2X Mode 4 analytical models have been implemented in Matlab[6]. The models are validated in this section by comparing their outcome with that obtained with a C-V2X Mode 4 simulator developed over Veins and presented in [10]. The results obtained with this simulator are used as benchmark since no other open-source C-V2X Mode 4 implementation is currently available, and to the authors' knowledge, no analytical models of the C-V2X Mode 4 communication performance have been reported in the literature. Veins integrates OMNET++ for wireless networking simulation with the open-source traffic simulation platform SUMO. The simulations conducted utilize realistic mobility of vehicles using the open source traffic simulator

---

[6] The implementation is released at: https://github.com/msepulcre/C-V2X

SUMO. SUMO models the mobility of vehicles using the Krauss car following model that maintains a safe distance between a vehicle and its vehicle in front, and selects the speed of vehicles so that vehicles can stop safely and avoid rear-end collisions. The mobility of vehicles has been generated for the highway scenario considered in this study, and following the parameters specified in Table III. The simulator implements the complete MAC of C-V2X Mode 4 including the sensing-based SPS scheme and the Winner+ B1 propagation model recommended by the European project METIS for D2D/V2V [17]. The physical layer performance is modelled through the link level LUTs presented in [15].

The comparison between the analytical models and the simulations is conducted considering that vehicles transmit packets at $\lambda$=10Hz with a transmission power $P_t$=20dBm and an MCS using QPSK and a coding rate of 0.7. This setting results in that each packet occupies 10 RBs, and there are hence 4 sub-channels per sub-frame. However, the models have been validated for other transmission power levels, different packet transmission frequencies, and an MCS using QPSK and a coding rate of 0.5 (2 sub-channels per sub-frame). Table III summarizes the main parameters considered for the validation, and that follow the 3GPP guidelines for the evaluation of C-V2X Mode 4 [18]. The simulations consider a highway of 5km with 4 lanes (2 lanes per driving direction) and vehicles moving at 70km/h. To avoid boundary effects, statistics are only taken from the vehicles located in the 2km around the center of the simulation scenario.

The accuracy of the proposed analytical models is estimated using the Mean Absolute Deviation (MAD) metric that quantifies the absolute difference between two vectors of $M$ elements, $m_s$ and $m_a$:

$$MAD[\%] = \frac{100}{M} \sum_{i=1}^{M} |m_s(i) - m_a(i)| \quad (35)$$

The MAD metric is here used to compare the PDR and the four possible transmission errors obtained through simulations and using the analytical model proposed. The comparison requires representing the results as vectors. The MAD metric represents then, as a percentage, the average difference between the results obtained analytically and through simulations. For example, a MAD equal to 1% means that on average the results obtained analytically and through simulations differ by 1%. The MAD metric numerically complements the visual comparison of the analytical and simulation results.

TABLE III. PARAMETERS

| Parameter | Values analyzed |
|---|---|
| Traffic density ($\beta$) | 0.1, 0.2, 0.3 veh/m |
| Avg. number of vehicles | 2000, 4000, 6000 |
| Max. speed of vehicles | 70 km/h |
| Highway length | 5km |
| Number of lanes | 4 (2 per direction) |
| Channel bandwidth | 10MHz |
| Transmission power ($P_t$) | 20, 23 |
| Packet tx frequency ($\lambda$) | 10, 20 Hz |
| Packet size ($B$) | 190 bytes |
| Sub-channels per sub-frame ($S$) | 2, 4 |
| RBs per sub-channel | 17 (2 sub-channels) <br> 12 (4 sub-channels) |
| Modulation and coding scheme | MCS 7 (QPSK 0.5, for 2 sub-channels) <br> MCS 9 (QPSK 0.7, for 4 sub-channels) |

## B. Validation

Fig. 8a compares the PDR curves obtained with the proposed analytical model (dashed lines) and with the C-V2X Mode 4 simulator (solid lines) for $P_t$=20dBm, $\lambda$=10Hz, 4 sub-channels per sub-frame, an MCS of QSPK with coding rate of 0.7, and different traffic densities. The figure clearly shows that the PDRs obtained with the proposed analytical model closely match those obtained by simulation. This trend is maintained irrespective of the traffic density and the resulting CBR. For example, a traffic density $\beta$ of 0.1veh/m resulted in an estimated CBR of approximately 0.23, while a traffic density of $\beta$=0.3veh/m resulted in a CBR[7] of 0.62. Fig. 8a shows that the analytical model is capable to provide an accurate PDR for low and high traffic densities, and hence channel load levels.

The proposed analytical models have also been evaluated for different transmission power levels. Fig. 8a depicted the PDR for $P_t$=20dBm, and Fig. 8b depicts it for $P_t$=23dBm. For the later, the analytical CBR ranged from 0.27 ($\beta$=0.1veh/m) to 0.69 ($\beta$=0.3 veh/m). Fig. 8b shows again that the PDRs obtained with the proposed analytical model closely match the ones obtained by simulation.

Another important parameter that influences the operation and performance of C-V2X Mode 4 is the number of packets transmitted per second per vehicle, $\lambda$. This parameter influences the number of sub-frames within the Selection Window, and the channel load and interference experienced by all vehicles. Fig. 8c shows the PDR obtained for $P_t$=20dBm and $\lambda$=20Hz for 3 traffic densities. The figure shows once more the close match between the PDRs obtained by simulation and using the proposed analytical models. For $\beta$=0.3 veh/m, the channel load was so high (analytical CBR of 0.85) that the proposed model slightly deviates from the simulation results (6.5% mean absolute deviation). However, it is important to consider that such high CBR levels would compromise the system's stability and scalability, and should hence be avoided using congestion control mechanisms. In fact, relevant studies recommend that the target CBR for V2X systems using IEEE 802.11p should be in the range of 0.6-0.7 [19] and ETSI recommends a default maximum CBR of 0.5 [20]. The 3GPP has not defined yet a target CBR for C-V2X.

The MCS influences the link level performance of C-V2X, the number of RBs that each packet occupies, and hence the number of sub-channels per sub-frame. The previous results were obtained with a MCS using QPSK and a coding rate of 0.7 (4 sub-channels per sub-frame). Fig. 8d shows the PDRs obtained with a MCS using QPSK and a coding rate of 0.5 (2 sub-channels per sub-frame). Fig. 8d demonstrates the validity of the presented analytical models for different MCS and number of sub-channels per sub-frame. The PDR is shown for traffic densities of 0.1, 0.2 and 0.3 veh/m that correspond to analytical CBR levels of 0.44, 0.74 and 0.86.

The accuracy of the proposed analytical models to calculate the probability of packet loss due to the different errors identified has also been evaluated. Fig. 9 depicts the probability of packet loss due to collisions as a function of the distance between transmitter and receiver for $P_t$=20dBm, $\lambda$=10Hz, 4 sub-channels per sub-frame, and different traffic densities. Fig. 9 shows that the proposed analytical model is also capable to accurately quantify this type of packet errors as its performance closely matches that obtained through simulations. The same accuracy is observed for different traffic densities. Fig. 9 shows that the probability of losing a packet due to collisions has a maximum around 350-400m. This is the distance at which the hidden-node problem causes higher degradation in this scenario.

Fig. 10 shows the probability of losing a packet due to the half-duplex effect, due to a received signal power below the sensing power threshold and due to the propagation. The probabilities are shown as a function of the distance between transmitter and receiver for $P_t$=20dBm, $\lambda$=10Hz and 4 sub-channels per sub-frame. These probabilities are independent of

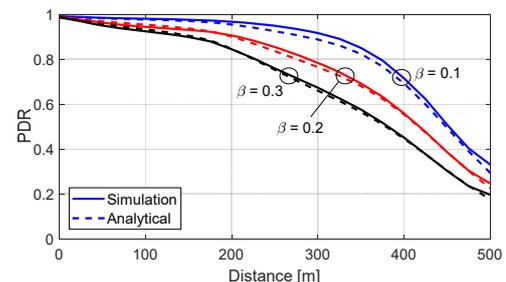
(a) $P_t$=20dBm, $\lambda$=10Hz, 4 sub-channels/sub-frame (QPSK 0.7).

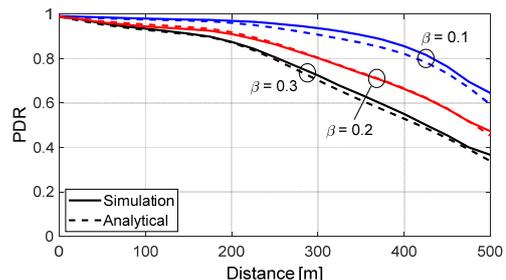
(b) $P_t$=23dBm, $\lambda$=10Hz, 4 sub-channels/sub-frame (QPSK 0.7).

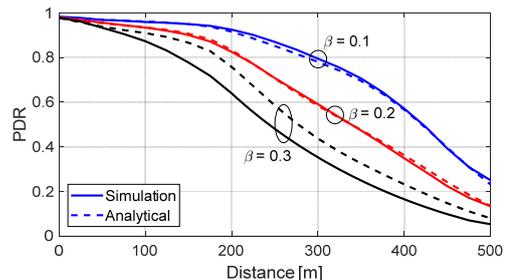
(c) $P_t$=20dBm, $\lambda$=20Hz, 4 sub-channels/sub-frame (QPSK 0.7).

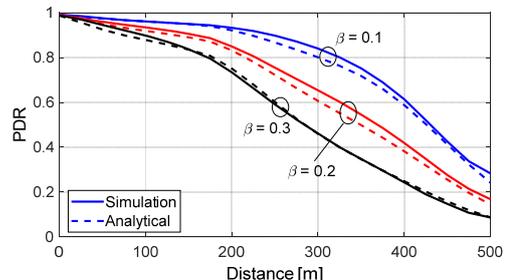
(d) $P_t$=20dBm, $\lambda$=10Hz, 2 sub-channels/sub-frame (QPSK 0.5)

Fig. 8. PDR as a function of the distance between transmitter and receiver for different traffic densities.

---

[7] CBR levels analytically estimated using eq. (34).

the traffic density. Fig. 10 shows again a good match between the values obtained by simulation and using the analytical models. The probability of losing a packet due to the half-duplex effect (Fig. 10a) depends on the duration of C-V2X sub-frames and the number of packets transmitted per second. However, it does not depend on the distance between transmitter and receiver or the traffic density. The probability of losing a packet due to propagation (Fig. 10a) is almost null at short distances, and has a maximum at around 450m to the transmitter. At higher distances, this probability decreases because most of the packets cannot even be detected due to a received signal power below the sensing power threshold. In fact, the probability of losing a packet because its received signal power is below the sensing power threshold increases as the distance to the transmitter increases (see Fig. 10b).

The accuracy of the proposed analytical models is analyzed in Tables IV, V and VI. The tables report the MAD metric for the PDR and the four possible transmission errors in C-V2X mode 4 under different conditions. The MAD metric is utilized to compare the results obtained analytically and through simulations. The MAD metric is shown for different transmission power levels, traffic densities, packet transmission frequencies, and number of sub-channels per sub-frame (or MCS). The tables also show in the last column the CBR level (analytically estimated) for each combination of parameters reported in the tables. The results obtained show that the PDR estimated analytically (using the models presented in this paper) differs on average by less than 2.5% compared to the PDR obtained through simulations in all scenarios where the CBR is below 0.8. In many cases, the deviation is smaller than 1%, which demonstrates the high accuracy that can be achieved with the proposed analytical

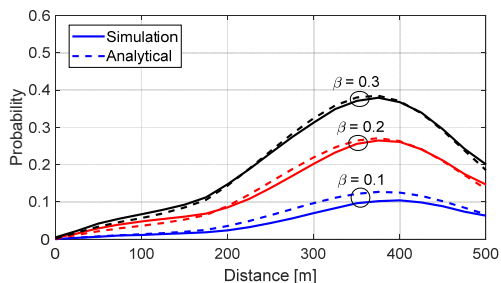

Fig. 9. Probability $\hat{\delta}_{COL}$ of packet loss due to collisions as a function of the distance between transmitter and receiver for $P_t$=20dBm, $\lambda$=10Hz, 4 sub-channels/sub-frame (QPSK 0.7) and different traffic densities.

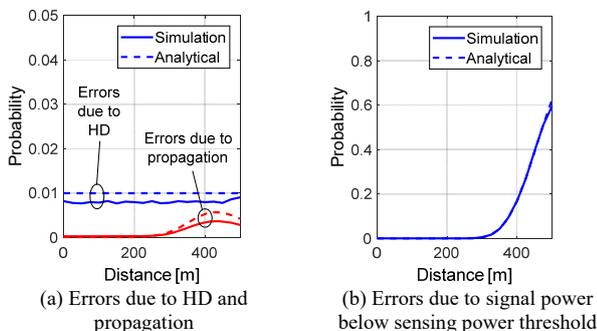

(a) Errors due to HD and propagation  
(b) Errors due to signal power below sensing power threshold

Fig. 10. Probability of losing a packet due to (a) HD and propagation effects, and due to a received signal power below sensing power threshold (b). $P_t$=20dBm, $\lambda$=10Hz and 4 sub-channels/sub-frame (QPSK 0.7).

models. The tables show that the type of error that has a higher contribution to the MAD of the PDF is actually the error due to packet collisions; this type of error was the most difficult to model due to the operation of C-V2X Mode 4 and its sensing-based SPS scheme.

TABLE IV. MAD FOR THE PDR AND THE DIFFERENT TYPES OF ERRORS. $\lambda$=10HZ AND 4 SUB-CHANNELS PER SUB-FRAME (QPSK 0.7)

| $P_t$ | $\beta$ | PDR | $\hat{\delta}_{HD}$ | $\hat{\delta}_{SEN}$ | $\hat{\delta}_{PRO}$ | $\hat{\delta}_{COL}$ | CBR |
|---|---|---|---|---|---|---|---|
|    | 0.1 | 1.60 | 0.19 | 0.21 | 0.07 | 1.25 | 0.23 |
| 20 | 0.2 | 0.91 | 0.16 | 0.18 | 0.07 | 0.83 | 0.44 |
|    | 0.3 | 0.92 | 0.14 | 0.20 | 0.07 | 0.84 | 0.62 |
|    | 0.1 | 1.95 | 0.21 | 0.14 | 0.05 | 1.59 | 0.27 |
| 23 | 0.2 | 0.52 | 0.17 | 0.13 | 0.05 | 0.65 | 0.51 |
|    | 0.3 | 1.24 | 0.18 | 0.15 | 0.05 | 0.94 | 0.69 |

TABLE V. MAD FOR THE PDR AND THE DIFFERENT TYPES OF ERRORS. $\lambda$=20HZ AND 4 SUB-CHANNELS PER SUB-FRAME (QPSK 0.7)

| $P_t$ | $\beta$ | PDR | $\hat{\delta}_{HD}$ | $\hat{\delta}_{SEN}$ | $\hat{\delta}_{PRO}$ | $\hat{\delta}_{COL}$ | CBR |
|---|---|---|---|---|---|---|---|
|    | 0.1 | 0.74 | 0.36 | 0.18 | 0.07 | 0.55 | 0.44 |
| 20 | 0.2 | 0.61 | 0.32 | 0.15 | 0.07 | 0.87 | 0.74 |
|    | 0.3 | 6.28 | 0.25 | 0.19 | 0.07 | 6.63 | 0.86 |

TABLE VI. MAD FOR THE PDR AND THE DIFFERENT TYPES OF ERRORS. $\lambda$=10HZ AND 2 SUB-CHANNELS PER SUB-FRAME (QPSK 0.5)

| $P_t$ | $\beta$ | PDR | $\hat{\delta}_{HD}$ | $\hat{\delta}_{SEN}$ | $\hat{\delta}_{PRO}$ | $\hat{\delta}_{COL}$ | CBR |
|---|---|---|---|---|---|---|---|
|    | 0.1 | 1.75 | 0.32 | 0.25 | 0.12 | 1.50 | 0.44 |
| 20 | 0.2 | 2.51 | 0.27 | 0.18 | 0.12 | 2.28 | 0.74 |
|    | 0.3 | 0.93 | 0.22 | 0.19 | 0.12 | 1.02 | 0.86 |

## VI. CONCLUSIONS

This paper has presented the first analytical models of the communication performance of C-V2X or LTE-V Mode 4. In particular, the paper has presented models of the average PDR as a function of the distance between transmitter and receiver, and of the four types of transmission errors that can be encountered in C-V2X Mode 4 communications. The models are validated in this paper for a wide range of transmission parameters (transmission power, packet transmission frequency, and MCS) and traffic densities. To do so, the paper compares the results obtained with the analytical models to those obtained with a C-V2X Mode 4 simulator implemented over the Veins platform. The conducted analysis has shown that the analytical models are capable to accurately model the C-V2X Mode 4 communications performance. In fact, the mean absolute deviation of the results obtained with the analytical models is generally below 2.5% compared with the results obtained by simulation. The analytical models hence represent a valuable tool for the community to evaluate and provide insights into the communications performance of C-V2X Mode 4 under a wide range of parameters.

This work paves the way for further studies and evolutions of C-V2X Mode 4. For example, the 3GPP standard does not specify concrete values for some of the parameters that define the operation and configuration of C-V2X Mode 4. In fact, ETSI is currently defining the default configuration of C-V2X Mode 4 parameters, and a detailed analysis of the optimum configuration of C-V2X Mode 4 is needed for the future deployment of C-V2X technologies. Also, different studies have highlighted possible inefficiencies of C-V2X Mode 4 to schedule the resources when the transmissions are not periodic. This is the case because of the semi-persistent nature of the scheduling scheme of C-V2X Mode 4 that results in a loss of efficiency if vehicles need to frequently reselect resources, or if they do not fully utilize the reserved resources.

## APPENDIX A

Eq. (5) expresses the PDR as a function of the different error probabilities (all of them between 0 and 1). Eq. (6) expresses the PDR as a function of the normalized probabilities so that their sum is always below than or equal to 1. If we substitute the normalized error probabilities in eq. (6) by their expressions in eq. (6.1)-(6.4) we obtain:

$$PDR(d_{t,r}) = 1 - \delta_{HD} - \left((1-\delta_{HD}) \cdot \delta_{SEN}(d_{t,r})\right) \\ - \left((1-\delta_{HD}) \cdot (1-\delta_{SEN}(d_{t,r})) \cdot \delta_{PRO}(d_{t,r})\right) \\ - \left((1-\delta_{HD}) \cdot (1-\delta_{SEN}(d_{t,r})) \cdot (1-\delta_{PRO}(d_{t,r})) \cdot \delta_{COL}(d_{t,r})\right) \quad (A.1)$$

We can then take out $(1-\delta_{HD})$ as a common factor to obtain:

$$PDR(d_{t,r}) = (1-\delta_{HD}) \cdot \big(1 - \delta_{SEN}(d_{t,r}) \\ - (1-\delta_{SEN}(d_{t,r})) \cdot \delta_{PRO}(d_{t,r}) \\ - (1-\delta_{SEN}(d_{t,r})) \cdot (1-\delta_{PRO}(d_{t,r})) \cdot \delta_{COL}(d_{t,r})\big) \quad (A.2)$$

Similarly, we can take out $(1-\delta_{SEN})$ as common factor of the right term of eq. (A.2) to obtain:

$$PDR(d_{t,r}) = (1-\delta_{HD}) \cdot (1-\delta_{SEN}(d_{t,r})) \cdot \\ \big(1 - \delta_{PRO}(d_{t,r}) - (1-\delta_{PRO}(d_{t,r})) \cdot \delta_{COL}(d_{t,r})\big) \quad (A.3)$$

We can then take out $(1-\delta_{COL})$ as common factor to obtain eq. (A.4), which is equal to eq. (5):

$$PDR(d_{t,r}) = (1-\delta_{HD}) \cdot (1-\delta_{SEN}(d_{t,r})) \\ \cdot (1-\delta_{PRO}(d_{t,r})) \cdot (1-\delta_{COL}(d_{t,r})) \quad (A.4)$$